\documentclass{article}
\usepackage{epsfig}
\usepackage{amsmath}
\usepackage{amsfonts}\tolerance=10000
\date{December 2009} 
 
\def\eqn{\begin{equation}} 
\def\eeqn{\end{equation}}
\def\arr{\begin{array}} 
\def\earr{\end{array}}
\def\eqna{\begin{eqnarray}} 
\def\eeqna{\end{eqnarray}} 

 \def\D{\Delta}

\def\D{\Delta}

\def\m{\mu}

\usepackage{graphicx}

\newcommand{\insertplot}[5]{\begin{figure}
 \hfill\hbox to 0.05in{\vbox to #5in{\vfill
 \inputplot{#1}{#4}{#5}}\hfill}
 \hfill\vspace{-.1in}
 \caption{#2}\label{#3}
 \end{figure}}
 \newcommand{\inputplot}[3]{
 \special{ps: plotfile #1}
}
\newcounter{fig}

\newcommand{\beq}{\begin{equation}}
\newcommand{\eeq}{\end{equation}}
\newcommand{\beqs}{\begin{eqnarray}}
\newcommand{\eeqs}{\end{eqnarray}}

\numberwithin{equation}{section}
\newcommand{\be}{\begin{equation}}
\newcommand{\ee}{\end{equation}}
\newcommand{\bea}{\begin{eqnarray}}
\newcommand{\eea}{\end{eqnarray}}

\usepackage{graphicx}

\begin{document}

\title{\bf On the Bekenstein-Hawking area law \\ for black objects
with conical singularities} 

\author{
{\large Carlos Herdeiro}$^{\dagger}$,
{\large Burkhard Kleihaus}$^{\ddagger}$, {\large Jutta Kunz}$^{\ddagger}$ 
and {\large Eugen Radu}$^{\ddagger}$  
\\ 
\\
$^{\dagger}${\small  Centro de F\'{\i}sica do Porto e Departamento de F\'{\i}sica da Faculdade de Ci\^encias da Universidade do Porto,}
\\ {\small Rua do Campo Alegre 687, 4169-007 Porto, Portugal} 
\\
$^{\ddagger}${\small  Institut f\"ur Physik, Universit\"at Oldenburg, Postfach 2503
D-26111 Oldenburg, Germany}
}
 
\maketitle 
 
\begin{abstract}
We argue that, when working with the appropriate set of thermodynamical variables, the Bekenstein-Hawking law still holds for
asymptotically flat black objects with conical singularities. The mass-energy which enters the first law of thermodynamics 
does not, however, coincide with the ADM mass; it differs from the latter by the energy associated with the conical singularity, as seen by an asymptotic, static observer. These statements are supported by a number of examples: the Bach-Weyl (double-Schwarzschild)  solution, its  dihole generalisation in Einstein-Maxwell theory and the five dimensional 
static black ring.
\end{abstract}

\section{ Introduction}

Conical singularities have a topological nature. They arise from identifications along the orbits of a $U(1)$ isometry, in the neighbourhood of a fixed point of this symmetry. It is therefore somewhat surprising that there are circumstances under which the local equations of general relativity demand the existence of conical singularities. 

One such circumstance is realised in non-extremal multi-black hole solutions. The long range interactions in these solutions are unable to provide equilibrium between the various black holes. The role of the conical singularities is to provide the force balance that allows the existence of such solutions in a non-linear theory. The basic example is the 
Bach-Weyl (BW) or double-Schwarzschild solution \cite{BW}, which describes two static black holes in four dimensional vacuum general relativity. This configuration belongs to the general class of 
static Ricci flat vacua with a $U(1)$ isometry found by Weyl \cite{Weyl}, and it was generalised to $N$ collinear Schwarzschild black holes by Israel and Khan \cite{IsraelKhan}. The occurrence of conical singularities in the BW solutions was first discussed by Einstein and Rosen \cite{Einstein:1936fp}; the generic solution must have these singularities, albeit their precise location is a matter of choice. Depending on the choice of periodicity for the coordinate along the $U(1)$ orbits, the conical singularity can be put either in between the two black holes - in which case it is interpreted as a \textit{strut} - or connecting either black hole to infinity - in which case it is interpreted as a \textit{string}. In order for the spacetime to be asymptotically flat (without any conical singularities at spatial infinity), we shall take the former viewpoint. The strut energy is then interpreted 
as the interaction energy between the black holes, while its pressure prevents 
the gravitational collapse of the system.

Another such circumstance is realised in unbalanced single black hole solutions. In more than four spacetime dimensions, there are black objects
with a topologically non-spherical horizon that, in vacuum, can only be balanced in a regular (on and outside the event horizon) geometry by introducing rotation. Their static version possesses conical singularities. The basic example is the 
static black ring \cite{Emparan:2001wn,Emparan:2001wk} in five spacetime dimensions. As before, the precise location of the conical singularity is a matter of choice. Demanding regularity at spatial infinity one finds a (spatially) two dimensional, compact, disk-like strut, which is located inside the non-trivial cycle of the ring, and that prevents this black object from collapsing. The black ring can then be used to construct vacuum multi-black objects supported by conical singularities, like 
the static black Saturn \cite{Elvang:2007rd}, 
the static bicycling black rings \cite{Elvang:2007hs}
and the static black di-rings \cite{Iguchi:2007is}. 
All these solutions may be made free of conical singularities by introducing rotation.

On general grounds, one expects  
the black hole solutions
with conical singularities to have well defined 
thermodynamics \cite{Gibbons:1979nf}. This is indeed the case for the BW two black holes configuration \cite{Costa:2000kf}. The new feature  introduced by the conical singularity is an extra contribution to the total tree level  Euclidean action of the system and consequently a new term in the first law of thermodynamics. But somewhat intriguingly, the results in \cite{Costa:2000kf} violate the Bekenstein-Hawking relation between area and entropy, for generic separation between the black holes. 

The Bekenstein-Hawking area law is a robust relation in general relativity, applying not only to black holes but also to cosmological horizons \cite{Gibbons:1977mu}. Moreover it does not apply only to regular solutions; there are known examples of horizons with conical singularities for which the area law still holds. In \cite{Aryal:1986sz} it was argued that it holds for a Schwarzschild black hole with a cosmic string passing through it, and in   \cite{Emparan:2001wk} it was argued that it holds for a static black ring with a disk-like conical deficit extending from the horizon to infinity. The area law loses its validity when one departs from general relativity \cite{Wald:1993nt}, but, as far as we are aware, there are no other known asymptotically flat solutions within general relativity which violate the area law.\footnote{The presence of NUT charge violates the area law \cite{Hawking:1998jf}, but solutions with NUT charge are not asymptotically flat.} 
 Thus, it becomes important to clarify the relation between area and entropy for the BW solution.

With this goal in mind, we analyse, in this paper, some basic thermodynamical 
quantities for several asymptotically flat solutions with conical singularities. 
We argue that, when working with the appropriate set of thermodynamical variables, 
the Bekenstein-Hawking entropy/area relation
 still holds, generically, in the presence of  conical singularities. 
 The mass-energy which enters the first law of thermodynamics does not, 
 however, coincide with the ADM mass; it differs from the latter by the energy associated with the conical singularity, as seen by an asymptotic, static observer. These statements are supported by the results we find for a number of different asymptotically flat solutions containing
conical singularities. These are the four dimensional BW solution and its 
black dihole generalisation in Einstein-Maxwell theory and  the five dimensional static black ring.
 
 This paper is organised as follows. In section 2 we discuss some general features
of a spacetime containing  conical singularities and present our proposal for the 
set of variables which enters the first law of thermodynamics.
Section 3 contains a discussion of the  thermodynamics
for multi-black holes systems: we consider both the BW solution and the dihole in Einstein-Maxwell theory.
We discuss the Emparan-Reall static black ring in section 4 
and give our conclusions and remarks in the final section. 

To simplify the general relations, we shall set $G_d=1$ for the Newton's constant in $d-$dimensions.

\section{ General remarks}
 
The configurations we shall consider herein are static and axially symmetric,  admitting $d-2$
orthogonal, commuting, non-null Killing vectors, where $d$ is the spacetime dimension.
Then, it follows that their line element may be written in the form
\begin{eqnarray}
\label{metric}
ds^2=-e^{2U_1(\rho,z)}dt^2+\sum_{i=2}^{d-2} e^{2U_i(\rho,z)}(dx^i)^2+e^{2\nu(\rho,z)}(d\rho^2+dz^2),
\end{eqnarray}
where   $0\leq \rho<\infty,$ $-\infty< z<\infty$. 
For $d=4$ this is the usual Weyl canonical form \cite{Weyl},
which has been generalised in \cite{Emparan:2001wk}
to dimensions higher than four.\footnote{Note that an
asymptotically flat spacetime can be described by this form for $d=4,5$ only.}

Typically, in this class of solutions, the $x^i$-coordinates are angular (azimuthal) variables. Denote one such coordinate by $\varphi$, with period  $\Delta \varphi$. We shall consider asymptotically flat black objects possessing a conical singularity at $\rho=0$
on a finite $z-$interval. To define a
conical singularity on a fixed point set of a $U(1)$ isometry, with orbits parameterised by $\varphi$, 
 one computes
the proper circumference $C$ of these orbits and their
proper radius $R$ and
defines
\begin{eqnarray}
\alpha&=&\frac{dC}{dR}\bigg|_{R=0}=\lim_{\rho\rightarrow 0}\frac{\sqrt{%
g_{\varphi\varphi}}\Delta\varphi}{\int_0^{\rho}\sqrt{g_{\rho\rho}}d\rho}%
=\lim_{\rho\rightarrow 0}\frac{\partial_{\rho}\sqrt{g_{\varphi\varphi}}%
\Delta\varphi}{\sqrt{g_{\rho\rho}}}.
\end{eqnarray} 
The presence of a conical
singularity is now expressed by means of
\begin{eqnarray}
\label{delta}
\delta&=&2\pi- \alpha,
\end{eqnarray}
such that $\delta>0$ ($\delta<0$) corresponds to a conical \textit{deficit} (\textit{excess}).

The physical interpretation of $\delta$ is that it provides a pressure which prevents 
the system from collapsing.
This can be proven as follows.
In the presence of a conical singularity,
the manifold ${\cal M}_{\alpha}$ factorizes as ${\cal M}_{\alpha}=\mathcal{C}_\alpha \times \Sigma$ (actually a semi-direct product to be precise), where $\mathcal{C}_\alpha$ is the two-dimensional conical surface $\rho-\varphi$
and $\Sigma$ is the remaining $(d-2)$-dimensional  surface, which may be seen as the world-volume of the conical defect. 
The Riemannian geometry in the presence of conical defects is involved, because of the singular curvature. A method to tackle this problem is based on representing the singular manifold ${\cal M}_{\alpha}$ as the limit of
a converging sequence of smooth manifolds, on which the standard Riemannian
formulae hold \cite{Fursaev:1995ef}.
The results therein show that
the Ricci scalar of ${\cal M}_{\alpha}$, ${}^{(\alpha)}R$, can be represented near $\Sigma$ in the following local form:
\begin{eqnarray}
\label{R-tot}
{}^{(\alpha)}R=R+2(2\pi -\alpha) \delta_{\Sigma},
\end{eqnarray}
where 
$R$ is the curvature computed in the standard way on the
smooth domain ${\cal M}_\alpha/\Sigma$ of ${\cal M}_\alpha$.
Here, 
$\delta_{\Sigma}$ is the Dirac $\delta$-function, with $\int_{{\cal M}_\alpha} f \delta_\Sigma=\int_{\Sigma} f$.

An important consequence of the above equations is the following formula for
the integral curvature on the total manifold ${\cal M}_\alpha$:
\begin{eqnarray}
\label{new}
\int_{{\cal M}_\alpha }{}^{(\alpha)}R=\int_{{\cal M}_\alpha /\Sigma}R+2(2\pi-\alpha) Area,
\end{eqnarray}
where $Area$ is the area of $\Sigma$, $i.e.$ the space-time area of the conical singularity's world-volume. The contribution of the conical singularity, the last term on the right hand side, was first derived by Regge \cite{Regge}. The results in \cite{Fursaev:1995ef} also show that the singular part of the Ricci tensor has components only in the $\rho-\varphi$ plane,
such that $R_i^j=0$ for the remaining components.
Then one can use the Einstein equations $G_{ij}=8\pi T_{ij}$ to define an `effective stress energy tensor' associated 
with the singularity.
It follows that the only non-vanishing components of $T_{i}^j$
are 
\begin{eqnarray}
\label{tik}
T_i^j=-\delta_{i}^j\frac{1}{8\pi }(2\pi-\alpha)\delta_{\Sigma},~~~{\rm with}~~(i,j)\neq (\rho,\varphi).
\end{eqnarray}
A direct consequence of this result is that the conical deficit/excess,
 $\delta=2\pi -\alpha$,
corresponds to 
the pressure exerted by the strut. This is found  by integrating the $T_z^z$-component over $\mathcal{C}_\alpha$
\begin{eqnarray}
\label{res1}
P=\int_{\mathcal{C}_\alpha}  T_z^z =-\frac{\delta}{8\pi } .
\end{eqnarray}
Another quantity of interest is the total energy associated with the strut as seen by a static
observer placed at infinity. This is found  by integrating the $T_t^t$-component (with the appropriate red-shift factor) over a $t=t_0={\rm constant}$ hyper-surface, 
\begin{eqnarray}
\label{Eint}
E_{int}=-\int_{t=t_0} T_t^t = \frac{\delta}{8\pi } \frac{Area}{\beta}=-P{\cal A},
\end{eqnarray}
 where we have defined
\begin{eqnarray}
{\cal A}\equiv \frac{Area} {\beta};
\end{eqnarray} 
$\beta=1/T_H$ is the periodicity of the Euclidean time and $T_H$ is the Hawking temperature. In practice we compute the space-time area $Area$ in the Euclidean section with periodic time.

As originally proposed in \cite{Gibbons:1979nf}, a simple way to approach the thermodynamics of a system with a
conical singularity is to use the path-integral formulation of quantum
gravity \cite{GibbonsHawking1}. 
While such a formulation is undoubtedly not the last word on this subject, at least in the semiclassical limit
it yields a relationship between gravitational entropy and other relevant thermodynamic quantities. Following this route, the first step is to evaluate the total tree level Euclidean action of the system.
The new feature introduced by the conical singularity is to add an extra contribution to
this quantity. This contribution can be 
evaluated\footnote{Here we assume that the matter fields are regular everywhere.} by using the relation (\ref{new}).
  Then the  total action is 
 \begin{eqnarray}
 \label{tot-action}
I=I_0-\frac{\delta}{8\pi}{\cal A}\beta, \label{action}
\end{eqnarray}
 where $I_0$ is the tree level action
found when neglecting the conical singularity. 
Then it is clear that the conical singularity
should also change the thermodynamics of the system; the first law of thermodynamics should contain a 
work term ${\cal T} dX$ originating from the conical singularity contribution.
Restricting to static, vacuum solutions, one should write
  \begin{eqnarray}
\label{firstlaw} 
d{\cal M}=T_H dS + {\cal T} d X,
  \end{eqnarray} 
  where $X$ is an extensive parameter which takes into account the presence of conical singularity and ${\cal T}$
 is the associated ``tension".  
 
  In a canonical ensemble, one keeps the Hawking temperature $T_H$ and the  extensive parameter $X$ fixed.
The free energy $F$ is 
 \begin{eqnarray}
\label{F} 
 F[T_H,X]=T_H I={\cal M}-T_H S.
  \end{eqnarray}
Then
the entropy $S$, mass ${\cal M}$ and tension ${\cal T}$ of the physical system are given by
\begin{eqnarray}
\label{r1} 
 S=-\frac{\partial F}{\partial T_H}\bigg|_{X},~~
~{\cal M}=F+T_HS,~~{\cal T}= \frac{\partial F}{\partial X}\bigg|_{T_H}.
 \end{eqnarray}

 At this stage one should decide which are the relevant thermodynamical parameters  ${\cal T},~X$.
 Previous work on the Bach-Weyl solution \cite{Costa:2000kf}, for instance,
 chose $X$
 to be the coordinate length $\Delta z$ of the strut and ${\cal T}$ to be the associated force exerted by the strut. Although $\Delta z$ is a coordinate quantity, this choice may be partly justified by the fact that the proper length $L$ of the strut is a monotonic function of $\Delta z$. But it is unclear, for instance, why it is the force, rather than the length, which is fixed in the first law \eqref{firstlaw}.  Moreover, this choice leads to an intriguing result that the entropy does not follow the area law, a consequence which does not change if the proper, rather than the coordinate, length is chosen, as we shall see below.

We suggest that the most natural set of variables is the one resulting from
(\ref{res1}), (\ref{Eint}) $i.e.$
$ {\cal T}= P=- {\delta}/{8\pi}$
for the ``tension" and $X={\cal A}$ for the conjugate extensive variable.
As we shall see in the examples below, this choice leads to a simple thermodynamical description.
In particular, the  Bekenstein-Hawking entropy/area relation holds for all cases we have considered.

Another argument that favours this choice of variables comes from considering the regularity of the Euclidean section. 
As mentioned in the introduction, by fixing the periodicity $\Delta \varphi$ one can change the spacetime region in which there are conical singularities. In particular, for the BW solution, one insures that the conical singularity $\delta$ vanishes either at infinity or in between the two black holes. Thus, for fixed black hole masses and distance between them, fixing $\delta$ is analogous to fixing the temperature $T_H$ in the Euclidean section: it determines the absence of conical singularities in some region of the Euclidean section. 
This is consistent with the fact that, for our choice of variables, $T_H$ and ${\cal T}=-\delta/8\pi$ should 
enter the first law on equal footing, cf. \eqref{firstlaw}.

We close this section by presenting a simple Smarr formula for the 
above set of variables.
This relation can be obtained from the first law \eqref{firstlaw} after taking into account scaling dimensions (see the similar approach in the black string literature \cite{Kol:2003if}). 
Here we perform a scaling transformation of the proper length $L$ of the strut, $L \to L+dL$,
which implies
\begin{eqnarray}
\label{SM1} 
{\cal M} \to {\cal M}\left(1+\frac{dL}{L}\right)^{d-3},~~
S\to S\left(1+\frac{dL}{L}\right)^{d-2},~~
{\cal A}\to {\cal A}\left(1+\frac{dL}{L}\right)^{d-3},
\end{eqnarray} 
and, therefore
\begin{eqnarray}
\label{SM2} 
\frac{d{\cal M}}{{\cal M}}=(d-3)\frac{dL}{L},~~
 \frac{dS }{S}=(d-2)\frac{dL}{L},~~
 \frac{d{\cal A}}{{\cal A}}=(d-3)\frac{dL}{L}.
\end{eqnarray} 
Replacing in the first law \eqref{firstlaw} we obtain the 
  Smarr formula
\begin{eqnarray}
\label{smarrform} 
 T_H S=\frac{(d-3)}{(d-2)} ({\cal M}-{\cal T A}).
\end{eqnarray}
Comparing with \eqref{action} and \eqref{F} we conclude that 
\begin{equation}
I_0=\frac{\beta}{d-2}\left(\mathcal{M}-\mathcal{T}\mathcal{A}\right), \end{equation}
which is verified by all vacuum solutions in this work.

\section{Two black hole systems in $d=4$}

\subsection{The $\mathbb{Z}_2$ symmetric Bach-Weyl solution}
In order to have a system of two black holes in thermodynamical equilibrium, the black holes must have the same Hawking temperature. In the case of the BW solution this means they must have the same mass. Then, a $z$ coordinate may be chosen, such that the solution is invariant under a discrete $\mathbb{Z}_2$ symmetry. The resulting line element is of the form (\ref{metric}), with $x^1=\varphi$ and in the conventions of Ref. \cite{Costa:2000kf},
\begin{eqnarray}
\label{metric2BHs1} 
&&e^{2U_1}=\frac{(r_1+r_2-\mu)}{(r_1+r_2+\mu)}\frac{(r_3+r_4-\mu)}{(r_3+r_4+\mu)},~~e^{2U_2}= \rho^2 e^{-2U_1} ,
\\
\nonumber
&&e^{2\nu}=e^{-2U_1}\left(\frac{\Delta z}{\Delta z+\mu}\right)^2
\left(\frac{(r_1+r_2)^2-\mu^2}{4r_1r_2}\right)
\left(\frac{(r_3+r_4)^2-\mu^2}{4r_3r_4}\right)
\left(\frac{(\Delta z+\mu) r_1+(\Delta z+2\mu)r_2-\mu r_4}{\Delta z~r_1+(\Delta z+\mu)r_2-\mu r_3}\right)^2,
\end{eqnarray}
where
\begin{eqnarray}
\label{Ri}
&&r_1=\sqrt{\rho^2+\left(z-\frac{\Delta z}{2}-\mu\right)^2},~~ r_2=\sqrt{\rho^2+\left(z-\frac{\Delta z}{2}\right)^2},~~ 
\\
\nonumber
&&r_3=\sqrt{\rho^2+\left(z+\frac{\Delta z}{2} \right)^2},~~ r_4=\sqrt{\rho^2+\left(z+\frac{\Delta z}{2} +\mu\right)^2}.
 \end{eqnarray}
The ADM mass of the spacetime (twice the individual black hole masses) is $\mu$ and $\Delta z$ is the coordinate distance between the two horizons along the $z$ axis. The event horizons are located at $\rho=0$, $-\Delta z/2 -\mu\leq z\leq -\Delta z/2$
and $ \Delta z/2 \leq z\leq  \Delta z/2+\mu$. The geodesic motion on the symmetry plane of this solution displays some unusual properties, as has been recently discussed in Ref. \cite{Coelho:2009gy}.

This system has a deficit angle along the section in between the black holes ,
$i.e.$ for $-\Delta z/2 \leq z\leq \Delta z/2$:
\begin{eqnarray}
\frac{\delta}{2\pi}=-\frac{\m^2}{(\D z+2\m)\D z}\ .
\end{eqnarray}
The event horizon area of each black hole and the corresponding Hawking temperature are  \cite{Costa:2000kf}:
\begin{eqnarray}
A_H= 4 \pi \mu^2 \frac{\Delta z+2\mu }{\Delta z+\mu},~~T_H=\frac{1}{4 \pi \mu}\frac{\Delta z+\mu}{\Delta z+2\mu}.
\end{eqnarray}
The proper length of the rod with conical singularity is
 \begin{eqnarray}
L=\Delta z \left(\frac{\bar\alpha+4}{\bar\alpha+2}\right)^2 E(m),
\end{eqnarray}
where $E(m)$ is the complete elliptic integral of the second kind
and
  \begin{eqnarray}
\bar \alpha=\frac{2\Delta z}{\mu},~~m=\left(\frac{\bar\alpha }{ \bar\alpha+4}\right)^2.  
 \end{eqnarray}
 The line element of the two dimensional surface $\Sigma$ spanned by the conical singularity for the BW solution reads 
   \begin{eqnarray}
ds^2=-\frac{\Delta z^2-4 z^2}{(\Delta z+2\mu)^2-4 z^2}dt^2
+\frac{\Delta z^2(\Delta z+2\mu)^2}{(\Delta z+\mu)^4}  \frac{(\Delta z+2\mu)^2-4 z^2}{\Delta z^2-4 z^2}dz^2.
 \end{eqnarray}
Therefore one finds 
 \begin{eqnarray}
{\cal A}=\frac{Area}{\beta}= \frac{(\Delta z)^2(\Delta z+2\mu)}{(\Delta z+ \mu)^2}, 
\qquad E_{int}= \frac{\delta}{8\pi}\mathcal{A}=-\frac{\mu^2\Delta z}{4(\Delta z+\mu)^2}.
\end{eqnarray}

\subsubsection{Thermodynamics}
 The first step is to move from the metric variables $\mu$, $\Delta z$
to those which enter the thermodynamical description in a canonical ensemble $T_H,\mathcal{A}$. This transformation is
 \begin{eqnarray}
 &&\mu=\frac{1}{8\pi T_H}
 \left (1+ \pi {\cal A} T_H-\sqrt{ \pi {\cal A} T_H(1+ \pi {\cal A}T_H) }
 \right )^{-1},
 ~~\Delta z= \frac{ {\cal A}}{2} \left(1+\frac{1+2\pi  {\cal A} T_H}{2 \sqrt{\pi {\cal A} T_H (1+ \pi {\cal A}  T_H)} }\right).
\end{eqnarray}
One can easily compute the action  (\ref{tot-action}) of the BW solution  and thus 
the free energy $F=T_H I$.  As expressed in terms of $T_H$ and ${\cal A}$, the free energy reads
\begin{eqnarray}
F[T_H,{\cal A}]=\frac{1}{16\pi T_H} -\frac{{\cal A}}{8} +\frac{1}{8\pi} \sqrt{\frac{\pi {\cal A}}{T_H}}\sqrt{ 1+ \pi {\cal A}  T_H }.
\end{eqnarray}
Therefore the entropy, mass and tension as computed from (\ref{r1})
are
\begin{eqnarray}
&&S=\frac{1}{16\pi T_H^2}+\frac{{\cal A}}{16 T_H \sqrt{\pi {\cal A}T_H(1+\pi {\cal A}T_H)}},
\\
&&{\cal M}=\frac{1}{8 \pi T_H}- \frac{{\cal A}}{8} +\frac{ {\cal A}(3+2 \pi {\cal A} T_H) }{16 \sqrt{\pi {\cal A}T_H(1+\pi {\cal A}T_H)}},
\\
&&-{\cal T}=\frac{1}{8}\left(1-\frac{1+2\pi {\cal A}T_H}{2 \sqrt{\pi {\cal A}T_H(1+\pi {\cal A}T_H)} }\right)=\frac{\delta}{8\pi},
\end{eqnarray}
the tension being just the pressure exerted by the strut, as expected.
One can easily see that the two black hole system is thermally unstable, $(\partial S/\partial T_H)|_{{\cal A}}<0$.

Moving back to the metric parameters $\mu$, $\Delta z$ one finds simple results:
\begin{eqnarray}
 S=2\frac{A_H}{4} ,
 ~~~{\cal M}=\mu\left(1+\frac{\mu\Delta z }{4(\mu+\Delta z)^2}\right)=M_{ADM}-E_{int}.
 \end{eqnarray} 
The first equation is the standard Bekenstein-Hawking area law, for a two black hole system. 
The second equation shows that the thermodynamical mass is the ADM mass minus the energy of the strut as seen by a static observer at infinity.
 In the limit of large separation between black holes, $\Delta z\to \infty$, one finds
\begin{eqnarray}
S=2\pi \mu^2 +\frac{2\pi \mu^3}{ \Delta z}+\dots,
~~~
\mathcal{M}=\mu +\frac{\mu^2}{4\Delta z}+\dots ,
\end{eqnarray}
which agree with the results in \cite{Costa:2000kf}, wherein $\Delta z$ has been taken as a thermodynamical parameter, instead of ${\cal A}$. Indeed it was observed therein that the area law was preserved to this order. Moreover, a quantitative matching of the entropy for a charged version of this solution with a string theory setup, was verified in  \cite{Costa:2000kf}, but only up to this order. We emphasise that for a finite $\Delta z$, however, the results there differ from what we have found here.

For completeness, we present also the results obtained when considering the proper length $L$, instead of the extensive variable  ${\cal A}$, in the thermodynamical description of the system.
A straightforward computation leads to the following expressions for mass, entropy and tension for this set of variables:
 \begin{eqnarray}
&&{\cal M}=\frac{1}{2\mu}
\left(
1+\frac{\bar\alpha}{(\bar\alpha+2)^2}+\frac{(\bar\alpha+4)}{(\bar\alpha+2)}\frac{F_1(\bar\alpha)}{F_2(\bar\alpha)}
\right)\neq \mu ,
\\
&&S=2 \pi \mu^2 \left(\frac{ \bar\alpha+4 }{ \bar\alpha+2 }\right)^2\frac{F_1(\bar\alpha)}{F_2(\bar\alpha)}\neq 2\frac{A_H}{4},
~~{\rm and}~~
{\cal T}=\frac{1}{F_2(\bar\alpha)},
\end{eqnarray}
where
\begin{eqnarray}
F_1(\bar\alpha)=(\bar\alpha+2)(\bar\alpha+4)E(m)-4(\bar\alpha+1) K(m),~~
F_2(\bar\alpha)= (\bar\alpha+4)^2 E(m)-4(\bar\alpha+2) K(m), 
\end{eqnarray}
 with $K(m)$
being the complete elliptic integral of the first kind. 
 As expected, the results in the limit $\Delta z\to \infty$ and $\mu$ 
 fixed  are in agreement with those in \cite{Costa:2000kf}, 
 where the chosen variable
 was $X=\Delta z$.  
The other limit of interest is  $\bar\alpha\to 0$ ($i.e.$  the two black holes
merge to form a single object). There one finds
\begin{eqnarray}
{\cal M}=\mu +\frac{\mu \bar\alpha^3}{128}+\dots,
~~
S=4\pi \mu^2-2\pi \mu^2 \bar\alpha+\dots,
~~
{\cal T}=\frac{1}{4\pi}-\frac{\bar\alpha}{8\pi}+\dots.
 \end{eqnarray}
 Besides the non-validity of the Bekenstein-Hawking area law, the fact that  ${\cal T} \to 1/4\pi$ in this limit, makes this description awkward.

\subsection{The dihole solution in Einstein-Maxwell theory}
By employing a suitable Harrison transformation 
(or a boost along the fifth dimension in the case of Kaluza-Klein theory), one can construct a
 a  generalisation of the BW two black holes solution to the Einstein-Maxwell-dilaton 
theory \cite{Costa:2000kf}. In this case, the  value of the $U(1)$ charge is the same for both black holes.
By using the same approach as in the vacuum case, one can prove that the entropy of these black holes 
is also one quarter of the total
horizon area, and that the relation $\mathcal{M}=M_{ADM}-E_{int}$ is still valid.
Moreover, the Smarr law for a charged  BW two black holes solution reads 
\begin{eqnarray}
\label{smarrform2} 
 T_H S=\frac{1}{2} ({\cal M}-{\cal T A}-\Phi Q_{tot}),
\end{eqnarray}
where $\Phi$ is the electrostatic potential difference between one of the horizons and infinity while
$Q_{tot}=2 Q$ represents the total electric charge of the configuration (with $Q$ the individual electric
 charge of a black hole \cite{Costa:2000kf}). As expected, both $\Phi$ and $Q_{tot}$ enter the first law in the usual way,
\begin{eqnarray}
d\mathcal{M}=T_HdS-\frac{\delta}{8\pi}d\mathcal{A} +\Phi dQ_{tot}\ . \label{newfirstlaw1}
\end{eqnarray}

Another interesting situation is provided by the black dihole in Einstein-Maxwell theory.
This configuration consists of a pair of non-extremal black holes at a finite distance, 
both with the same mass, and with charges of the
same magnitude but opposite sign (thus the total U(1) charge vanishes).
Since such solutions carry an electric (or magnetic) dipole moment, they are referred
to as \textit{diholes}.
These generalisations of the BW system
have been studied in \cite{Emparan:2001bb},
together with solutions to the $U(1)^4$ theories that
arise from compactified string/M-theory.

The Einstein-Maxwell dihole solution can also be written in the form (\ref{metric}),
with $x^2=\varphi$  (and $\Delta\varphi=2\pi$).
 The metric functions are then given by the following expressions
(here we follow the notations in \cite{Emparan:2001bb})
\begin{equation}
\label{f}
e^{2U_1}={\mathcal{A}^2-\mathcal{B}^2+\mathcal{C}^2\over
(\mathcal{A}+\mathcal{B})^2},~~~~e^{2U_2}=\rho^2 e^{-2U_1},~~
e^{2\nu}=e^{-2U_1}{\mathcal{A}^2-\mathcal{B}^2+\mathcal{C}^2
\over K_0R_+R_-r_+r_-},~
\end{equation}
where $\mathcal{A}$, $\mathcal{B}$ and $\mathcal{C}$ are given by
\begin{eqnarray} 
\nonumber
\mathcal{A}&=& -(R_+ -R_-)(r_+-r_-)[(\kappa_+^2 +
\kappa_-^2)(m^4+\kappa_+^2\kappa_-^2)-4 m^2\kappa_+^2\kappa_-^2]\cr
&&-2(\kappa_+^2-\kappa_-^2)(R_+R_-+r_+r_-)[m^4-(\kappa_+\kappa_-)^2] 
 +(\kappa_+^2-\kappa_-^2)(R_++R_-)(r_++r_-)[m^4+(\kappa_+\kappa_-)^2]\,,
\\
\mathcal{B}&=&
4m(\kappa_+^2-\kappa_-^2)\kappa_+\kappa_-[(R_++R_-+r_++r_-
)\kappa_+\kappa_-
-(R_++R_--r_+-r_-)m^2]\,,
\\
\nonumber
\mathcal{C}&=&
4kq\kappa_+\kappa_-[(R_+-R_-+r_+-r_-)(\kappa_+^2-m^2)\kappa_--(R_+-R_--
r_++r_-)(\kappa_-^2-m^2)\kappa_+],
\end{eqnarray}
and
\begin{equation}
R_\pm\equiv\sqrt{\rho^2+(z\pm(\kappa_++\kappa_-))^2}\,,\qquad
r_\pm\equiv\sqrt{\rho^2+(z\pm(\kappa_+-\kappa_-))^2}\, .
\label{rpm}
\end{equation}
The constants  $k$, $q$ and $K_0$ in the above expressions are 
\begin{equation}
 k=\sqrt{\kappa_+^2+\kappa_-^2-m^2}, ~~ q=\frac{\sqrt{(\kappa_+^2-m^2)(m^2-\kappa_-^2)}}{k},~ 
~~K_0 =(8\kappa_+^2\kappa_-^2(\kappa_+^2-\kappa_-^2))^2.
\end{equation}
Restricting ourselves to an electric configuration, the $U(1)$ potential of the dihole solution is
\begin{equation}
A=A_tdt,~~~{\rm with}~~A_t=-{\mathcal{C}\over \mathcal{A}+\mathcal{B}}~,
\end{equation}
(note that we choose a gauge such that $A_t$ vanishes at infinity).

This solution is written in terms of three parameters 
$\kappa_+,\kappa_-$ and $m$, which are restricted
to satisfy  $\kappa_-\leq m<\kappa_+$.
The mass of this system as computed at infinity is $ 2m$. 
Also, $\kappa_+$ controls the separation
between the black holes, while $\kappa_-$ is a `non-extremality'
parameter. 
The vacuum BW two black holes solution is recovered for $\kappa_-=m=\mu/2$, 
$\kappa_+ =(\mu+\Delta z)/2$, in which case the electric potential vanishes.

The black hole horizons lie on the axis $\rho=0$ at $-\kappa_-\leq
z\pm\kappa_+\leq\kappa_-$. 
The event horizon area  and Hawking temperature of each black hole are given by
\beq
A_H={4\pi\over\kappa_+}{(\kappa_++m)^2(\kappa_-
+m)^2\over \kappa_++\kappa_-}, ~~~
T_H=\frac{\kappa_+\kappa_-(\kappa_++\kappa_-
)}{2\pi (\kappa_++m)^2(\kappa_-+m)^2}.
\eeq
Similarly to the vacuum case, this system has a deficit angle along the section between black holes 
($i.e.$ for $-\kappa_+ + \kappa_-\leq z\leq \kappa_+ -\kappa_-$),
\beq
\delta =-2\pi{m^2(\kappa_+^2-m^2)+\kappa_+^2(m^2-\kappa_-^2)
\over(\kappa_+^2-m^2)^2}\,.
 \eeq
As discussed in  \cite{Emparan:2001bb}, the charge of each black hole is computed by employing
Gauss' law $Q={1\over 4\pi}\int_{S^2}*F$, taking the (topological)
sphere $S^2$ to
be a surface at constant small $\rho$ closely surrounding the black hole
horizon. Thus one finds the black hole electric
charge 
\beq
\label{Qe}
Q={\sqrt{(\kappa_+^2-m^2)(
m^2-\kappa_-^2)}\over \kappa_+-m}\, .
\eeq
The value $\Phi$ of the electric potential on the horizon is 
\beq
\Phi=\sqrt{(\kappa_+-m)(m-\kappa_-)\over(\kappa_++m)(m+\kappa_-)}.
\eeq
Both $Q$ and $\Phi$ change sign from one black hole to the other, but
the product $Q\Phi$ has the same sign for both.

A straightforward computation leads to the following expression for ${\cal A}=Area/\beta$:
\begin{eqnarray}
{\cal A}=\frac{2(\kappa_+^2-m^2)^2}{\kappa_+^2(\kappa_+ + \kappa_-)}.
\end{eqnarray}
Restricting to a canonical ensemble, the expression of the free energy $F=T_H I$ is 
\begin{eqnarray}
F=2m +\frac{1}{2}(\kappa_+-3\kappa_-)-\frac{(\kappa_+^2-m^2)^2}{2\kappa_+^2(\kappa_++\kappa_-)}.
\end{eqnarray} 
The thermodynamical description of the system contains, in this case, one extra parameter $Q$, which is 
associated with the contribution of the electromagnetic field\footnote{This corresponds to fixing the electric charge on a black hole. 
As remarked in \cite{Emparan:2001bb}, one should, more rigourously, define another thermodynamical ensemble that fixes, 
$e.g.$, the electric field (which is conjugate to the electric dipole moment). 
However we could not find a proper thermodynamical description in that case.}, 
the first law of thermodynamics being
\begin{eqnarray}
d{\cal M}=T_H dS+{\cal T} d{\cal A}+V dQ, 
\end{eqnarray}
with $Q$ the electric charge of  each  black hole, as given by (\ref{Qe}), and $V$
the associated potential.

Unfortunately, for the dihole solution it is not possible to express in a simple way the parameters
$\kappa_+$, $\kappa_-$ and $m$ in terms of ${\cal A},~T_H$ and $Q$.
Thus one has to determine first the expression of $d\kappa_+,d\kappa_-$ and $dM$ in terms of $d{\cal A}$, $dT_H$ and $dQ$.
However, a straightforward computation shows that the entropy is still one quarter of the total horizon area,
$S=2 A_H/4$ while the ``tension" associated to ${\cal A}$
is ${\cal T}=-\delta/8 \pi $.
The electric potential which enters the thermodynamics is 
\begin{eqnarray}
V=\frac{\partial F}{\partial Q}\bigg|_{T_H,{\cal A}}=2 \Phi.
\end{eqnarray}
For the thermodynamical mass one finds
\begin{eqnarray}
{\cal M}=2m +{\cal T A}=M_{ADM}-E_{int},
\end{eqnarray}
$i.e.$ apart from the ADM mass it contains again  the
total energy associated with the strut as seen by a static
observer placed at infinity.

\section{The $d=5$ static black ring}
This solution is of the form  (\ref{metric}), with
the angular coordinates $x^2=\psi$, $x^3=\varphi$, where $0\leq \psi,\varphi \leq 2 \pi$ in the asymptotic region. 
The  metric functions of the static black ring  are given  by \cite{Emparan:2001wk}, \cite{Harmark:2004rm}
\begin{eqnarray}
\nonumber
&&e^{2U_1} =\frac{R_2+\xi_2}{R_1+\xi_1},~~
e^{2\nu}= \frac{(R_1+\xi_1+R_2-\xi_2)((1-c)R_1+(1+c)R_2+2c R_3)}{8(1+c )R_1R_2R_3},
\\
\label{BR5d}
&&e^{2U_2} =  \frac{(R_2-\xi_2)(R_3+\xi_3)}{R_1-\xi_1},~~e^{2U_3} =R_3-\xi_3~,
\end{eqnarray}
  where
\begin{eqnarray}
\label{rel1}
\xi_i=z-z_i,~~
R_i=\sqrt{\rho^2+\xi_i^2}
~~~~
{\rm and~}~~~~
z_1=-a,~~z_2=a,~~z_3= b,
\end{eqnarray}
$a$ and $ b$ being two positive constants, with $c=a/b<1$. 
One can see that, at $\rho=0$, $e^{2U_2}$ vanishes for $ -\infty\leq z \leq -a $
and $ a \leq z  \leq b$, 
while $e^{2U_3}$ is zero for $ b \leq z \leq \infty$.
The event horizon is located at $\rho=0$, $ -a \leq z\leq a$.
Thus, since the orbits of $\psi$ shrink  to zero at $-a$ and $a$ while 
the orbits of  $\varphi$ do not shrink to zero size anywhere (for $ -a \leq z\leq a$), the topology of the horizon 
is $S^2\times S^{1}$.
 Other properties of these solutions are discussed in \cite{Emparan:2001wk}, \cite{Harmark:2004rm}.
 
 The mass $M$ as computed at infinity, the event horizon area $A_H$ and the Hawking temperature $T_H$ of the static black ring are:
\begin{eqnarray}
M =\frac{3\pi a}{2},~~
A_H =16 \sqrt{2} \pi^2 \frac{a^2}{\sqrt{ a+b} },~~
T_H=\frac{1}{4 \pi a}\sqrt{\frac{a+b}{2}}.
\end{eqnarray}
Although this solution is asymptotically flat\footnote{The $d=5$ static black ring solution in
  \cite{Emparan:2001wk} admits an alternative interpretation as 
  a ring sitting on the rim of a membrane that extends to infinity (this is found by requiring 
that the periodicity
of $\psi$ is $2\pi$ on the finite $\psi$-rod).
However, the asymptotic metric is a deficit membrane in this case.
The thermodynamics of this solution is discussed in  \cite{Emparan:2001wk}, wherein it is shown 
that the relation  $S=A_H/4$ is satisfied.}, it contains a conical excess angle $\delta$
for the  finite $\psi$-rod ($\rho=0$ and $a\leq z\leq b$)
\begin{eqnarray}
\delta=2\pi \left(1-\sqrt{\frac{b+a}{b-a}}\right)<0.
\end{eqnarray}
 The parameter $\cal A$ and  the proper length
 of the finite $\psi$-rod are
 \begin{eqnarray}
\label{rel11}
{\cal A}= \frac{2\pi (b-a)^{3/2}}{\sqrt{a+b}} ,~~L=\sqrt{2}\sqrt{b-a}~E\left(\frac{b-a}{b+a}\right),
\end{eqnarray}
$E(x)$ being again the complete elliptic integral of the second kind.

 A straightforward computation leads to simple (and unexpected) results for the entropy, thermodynamical mass and tension:
\begin{eqnarray}
\label{rel5}
S=\frac{A_H}{4},~~{\cal M}=\frac{\pi}{2}\left(2a +b- \frac{(b-a)^{3/2}}{\sqrt{b+a}}  \right)~~{\rm and}~~{\cal T}=-\frac{\delta}{8\pi}.
\end{eqnarray}
Again, the Bekenstein-Hawking area law is obeyed and the thermodynamical mass is $\mathcal{M}=M_{ADM}-E_{int}$, thus the ADM mass
minus the total energy associated with the strut as seen by a static
observer placed at infinity. 
Also, as expected, a static black ring is thermally unstable, $(\partial S/\partial T_H)|_{{\cal A}}<0$.

\section{Further remarks}
 One of the deepest concepts in gravitational physics is the relation between thermodynamical entropy and the event horizon area, described by the Bekenstein-Hawking formula. A natural question is then: how universal is this relation?  It is known that the area formula is corrected by quantum effects. This is manifest in higher derivative gravity \cite{Wald:1993nt}. A NUT charge, which changes the asymptotic structure of spacetime, also leads to a modification of the area law \cite{Hawking:1998jf}, \cite{Astefanesei:2004kn}. We may ask if the interaction 
 between two black holes, within general relativity and in an asymptotically flat configuration, may also violate this law.
 
This question may be answered, in a more general setup, by considering  black hole solutions of general relativity possessing conical singularities in the bulk. Intriguingly, some results in the literature suggest that  the entropy/area relationship can be violated by such black objects. The main purpose of this work was to present arguments  that, when working with the appropriate set of thermodynamical variables, the Bekenstein-Hawking law still holds for asymptotically flat black objects with conical singularities. The appropriate variables are the conical singularity $\delta$ and $\mathcal{A}$, the spacetime area that it spans (computed in the Euclidean section) multiplied by the Hawking temperature. All these quantities are invariants with a precise physical meaning. The first law then reads
\begin{eqnarray}
d\mathcal{M}=T_HdS-\frac{\delta}{8\pi}d\mathcal{A} 
+\sum_i {\cal  Z}_i dY_i ~,
  \label{newfirstlaw}
\end{eqnarray}
where $\sum_i {\cal Z}_i dY_i$ are other work terms which enter the usual thermodynamical
description of the system ($e.g.$ associated with rotation, the presence of a U(1) charge etc).

The mass-energy which enters the first law of thermodynamics, $\mathcal{M}$, that unfolds from this description does not coincide with the ADM mass; both differ by the energy associated 
with the conical singularity, as seen by an asymptotic static observer $E_{int}$
\begin{eqnarray}
\mathcal{M}=M_{ADM}-E_{int} \ . \end{eqnarray}
This result has been exhibited for different solutions in $d=4,~5$
spacetime dimensions. Technically, observe that the difference between computing $\mathcal{A}$ and the proper spatial size of the conical defect is that the redshift factor is considered in the volume element. This leads to considerable simplifications.

There is an important point for our analysis, which we have not yet addressed and that relates to the matter source that supports the conical excess in our solutions. Clearly, the solutions discussed herein,  which have (naked) conical singularities, should not be faced as vacuum (or electrovacuum) solutions. 
There is a matter source (the strut) which supports the conical singularity, as it is clear from eq. (\ref{tik}).
Strictly speaking, if this source is not taken into account, the Euclidean solution is not a saddle point of the functional integral. Then, one might conclude that there should be an extra contribution, from the matter action, to (\ref{tot-action}) which would modify the results we have presented. We shall now argue that our results are robust to properly taking into account the matter action. One starts by noticing that 
the energy-momentum tensor (\ref{tik}) results from a matter Lagrangean
\begin{eqnarray}
\label{lagr-con-sing} 
\mathcal{L}_m=-\frac{(2\pi -\alpha)}{8 \pi }\delta_{\Sigma}.
\end{eqnarray}
 This contribution to  the total tree level action of the system (including boundary term computed with respect to a reference background)
 \begin{equation} I=-\int\left(\frac{R}{16\pi}+\mathcal{L}_m\right)-\frac{1}{8\pi}\oint(K-K_0), \end{equation}
 gives $ Area (\delta/8\pi)=\mathcal{A}\beta (\delta /8\pi)$, cancelling the second term in (\ref{tot-action}) and therefore the contribution of the conical singularity. If one takes into account this matter contribution, however, one must for consistency consider a reference background with an equivalent source. By equivalent we mean a conical singularity with the same spacetime area $Area$ and conical deficit $\delta$. 
 The evaluation of the background action is a straightforward generalization
 of the computation in  \cite{Hawking:1995fd} for a piece of cosmic string of lenght $L$,
 the result being a contribution to the action of $-Area (\delta /8\pi)$. As the final result, one finds again (\ref{tot-action}) for the total action of the system, and learns the lesson that neglecting the matter contribution which is supporting the conical singularity effectively gives the correct action. Our approach is also consistent with that, for instance, in \cite{Emparan:2001wk}, wherein the static ring is considered with a conical deficit membrane extending to infinity. The authors also obtain the area law for the entropy, using Hamiltonian methods, by taking a deficit membrane spacetime as the reference background.

The generality of the results herein could be confirmed by considering various other examples. Obvious candidates are more involved static black objects, such as the $d=4$ Israel-Khan multi black holes, the $d=5$ multi-Tangherlini \cite{Tan:2003jz},  the $d=5$ black Saturn  or the
$d>5$ black holes with $S^2\times S^{d-4}$ topology of the horizon in \cite{Kleihaus:2009wh}.
It would also be interesting to compute the entropy of black objects with conical singularities in a cosmological background (however, a major obstacle in that case is the absence of exact solutions). Including rotation (but keeping conical singularities) should also be instructive. Perhaps the simplest case to consider is the unbalanced $d=5$ black ring 
rotating in a single plane.  There a new global charge appears, the angular momentum $J$, and the Smarr relation (\ref{smarrform}) becomes
\begin{eqnarray}
\label{smarrform1} 
 T_H S=\frac{2}{3} ({\cal M}-{\cal T A})-\Omega_H J,
\end{eqnarray}
with $\Omega_H$ being the event horizon angular velocity.
The equilibrium condition for a general rotating black ring solution
corresponds to ${\cal T}=0$. One may also consider the double Kerr solution (see e.g. \cite{Herdeiro:2008kq}; in the co-rotating case a discussion of equilibrium thermodynamics may be considered, as in Ref. \cite{Costa:2009wj}) or the double Myers-Perry solution \cite{Herdeiro:2008en}. Of course a more challenging problem is to give a general argument for \eqref{newfirstlaw}, without relying on specific examples.

As an avenue for future research, it would be interesting to study
the relation between the approach in this work and the quasi-local formalism.
For example, the recent paper \cite{Astefanesei:2009mc} uses the conservation of the renormalized boundary stress-energy
tensor within this formalism to 
 obtain the equilibrium condition for a  black ring solution. 
 Note, however, that the results in \cite{Astefanesei:2009mc} were found for the case of a conical singularity
 extending to infinity.

\section*{Acknowledgements}
C.H. would like to thank C. Rebelo and J. Santos for useful discussions.
C.H. is supported by a Ci\^encia 2007
research contract.  B.K. gratefully acknowledges support by the DFG. 
The work of E.R. was supported by a fellowship from the Alexander von Humboldt Foundation.  This work has been further supported
by the FCT Grant No. CERN/FP/83508/2008.

 \begin{small}
 
 \end{small}

\end{document}